\documentclass[conference]{IEEEtran}
\IEEEoverridecommandlockouts
\usepackage{cite}
\usepackage{amsmath,amssymb,amsfonts}
\usepackage{algorithmic}
\usepackage{algorithm}
\usepackage{graphicx}
\usepackage{textcomp}
\usepackage{xcolor}
\usepackage{booktabs}
\usepackage{url}
\usepackage{listings}
\usepackage{fancybox}
\usepackage{keyval}
\usepackage{stmaryrd} % \llbracket / \rrbracket
\def\BibTeX{{\rm B\kern-.05em{\sc i\kern-.025em b}\kern-.08em
    T\kern-.1667em\lower.7ex\hbox{E}\kern-.125emX}}

\DeclareFontShape{OT1}{ptm}{m}{scit}{<->ssub * ptm/m/sc}{}

\newif\ifcommentson
\commentsontrue

\newcommand{\ourtool}{\textsc{ConVer}}

\newcommand{\CMT}[1]{\hfill$\triangleright$ \textit{#1}}

\makeatletter
\define@key{dialogbox}{title}{\def\dialogbox@title{#1}}
\newsavebox{\dialogbox@box}
\newenvironment{dialogbox}[1][]{%
	\def\dialogbox@title{}%
	\setkeys{dialogbox}{#1}%
	\begin{lrbox}{\dialogbox@box}%
		\begin{minipage}{0.93\linewidth}%
			\vskip 6pt%
			\ifx\dialogbox@title\empty\else
				\colorbox{black!85}{%
					\makebox[\dimexpr\linewidth-2\fboxsep][l]{%
						\textcolor{white}{\bfseries\hspace{4pt}\dialogbox@title}%
					}%
				}%
				\vskip 1pt\leavevmode\\%
			\fi
			\ignorespaces
			}{%
			\vskip 6pt%
		\end{minipage}%
	\end{lrbox}%
	\par\vskip\medskipamount\noindent\shadowbox{\usebox{\dialogbox@box}}\par\vskip\medskipamount
}
\makeatother

\begin{document}

\title{\textsc{ConVer}: Using Contracts and Loop Invariant Synthesis for Scalable Formal Software Verification\\
\thanks{\textsuperscript{*}These authors contributed equally to this work.}}

\author{
\IEEEauthorblockN{Muhammad A. A. Pirzada\textsuperscript{*}}
\IEEEauthorblockA{\textit{The University of Manchester}\\
Manchester, UK\\
0009-0005-2440-7547}
\and
\IEEEauthorblockN{Weiqi Wang\textsuperscript{*}}
\IEEEauthorblockA{\textit{The University of Manchester}\\
Manchester, UK\\
0009-0005-8247-5994}
\and
\IEEEauthorblockN{Yiannis Charalambous\textsuperscript{*}}
\IEEEauthorblockA{\textit{The University of Manchester}\\
Manchester, UK\\
0009-0000-5755-5099}
\and[\hfill\mbox{}\\\mbox{}\hfill]
\IEEEauthorblockN{Konstantin Korovin}
\IEEEauthorblockA{\textit{The University of Manchester}\\
Manchester, UK\\
0000-0002-0740-621X}
\and
\IEEEauthorblockN{Lucas C. Cordeiro}
\IEEEauthorblockA{\textit{The University of Manchester}\\
Manchester, UK\\
0000-0002-6235-4272}
}

\maketitle

\begin{abstract}
Formal verification of large C programs is impeded by state-space explosion: Bounded Model Checking (BMC) tools must encode the entire state space up to the predetermined bound by unrolling all nested constructs. We present \ourtool{}, a top-down compositional verification tool. Given a C program with a top-level assertion, \ourtool{} decomposes verification top-down: it uses a large language model (LLM) to synthesise function contracts from the system property, then alternates system-level and function-level checks in a CEGAR-CEGIS loop, refining contracts whenever a check fails via SMART ICE learning. We evaluate \ourtool{} on four benchmark suites of increasing difficulty and against other state-of-the-art (SOTA) tools. On the Frama-C benchmark of 45~simple C programs, \ourtool{} achieves 82--96\% verification success across three LLM backends, with 93--95\% of converged programs requiring only a single CEGAR-CEGIS iteration. On the X.509 parser benchmark (6~programs) and LF2C-Simple suite (17~programs), \ourtool{} achieves 33--50\% and 82--88\% success respectively. On the VerifyThis suite of 11~recursive and loop-intensive programs, the Pre-Abstraction strategy achieves 55--64\% success.
In addition, we present \texttt{ESBMC-LF} a preprocessor tool that converts LF models to C while preserving the properties of the LF files, enabling \ourtool{} to verify them. We transpile the LF Verifier Benchmarks using \texttt{ESBMC-LF} to C; we denote those LF-Hard. We show that \ourtool{} successfully verifies 67\% of LF-Hard benchmarks overall.
\end{abstract}


%%%%%%%%%%%%%%%%%%%%%%%%%%%%%%%%%%%%%%%%%%%%%%%%%%%%%%%%%%%%%
\section{Introduction}
\label{sec:introduction}
%%%%%%%%%%%%%%%%%%%%%%%%%%%%%%%%%%%%%%%%%%%%%%%%%%%%%%%%%%%%%

Verifying the correctness of large C programs is one of the central challenges in software engineering and formal verification. BMC tools such as the Efficient SMT-based Context-Bounded Model Checker (ESBMC)~\cite{wu2025esbmc} and C Bounded Model Checker (CBMC)~\cite{cbmc-s2n} are highly effective on self-contained functions and small-scale models, but their complexity grows rapidly with program size: every function call adds to the state space that the underlying SMT solver must explore, and every loop requires unrolling up to a max bound. These contribute to the well-known state explosion problem in software model checking~\cite{clarke2011model}. The standard method of writing function contracts by hand is expert-intensive and poorly adopted in practice; even well-resourced projects such as AWS's \texttt{s2n-tls}~\cite{cbmc-s2n} require dedicated annotation effort from verification engineers. The result is that formal verification remains largely inaccessible for programs exceeding a few hundred lines.

\ourtool{} addresses this gap by combining two complementary ideas. First, it adopts a \emph{top-down property decomposition} strategy: rather than starting from individual function specifications and building upward, it begins with an abstract system-level assertion in \texttt{main()} and derives function contracts sufficient to prove that property. This top-down view minimises the specification burden as contracts need to only capture what the system property requires, not every semantic property of the underlying function. Second, \ourtool{} automates contract synthesis and refinement with Large Language Models (LLMs). An LLM proposes initial contracts; a software verification tool then checks them at the system level (by replacing calls with contract stubs) and at the function level (by verifying each implementation against its contract). Failures are then fed back to the LLM as structured counterexamples through a Counterexample Guided Abstraction Refinement (CEGAR)~\cite{clarke2000counterexample} loop augmented with SMART Implication Counterexamples (ICE) learning \cite{garg2014ice}, which classifies each counterexample into positive, negative, and implication examples to guide targeted refinement. \ourtool{} uses ESBMC for software verification.

% \paragraph{Motivating Example.}
% Figure~\ref{fig:architecture} illustrates the overall \ourtool{} pipeline. As a concrete illustration, consider a small alarm controller (\texttt{Alarm.c}) (see Appendix~\ref{sec:app:alarm_example} for more information) with the property ``the machine stops within one second of a fault''. \ourtool{} first extracts the assertion from \texttt{main()}, prompts the LLM to derive contracts for reaction functions, replaces their implementations with contract stubs, and ESBMC would check whether the property holds under those abstractions. If the system-level check passes, each function is then checked against its own contract individually. A full walkthrough appears in Section~\ref{subsec:alarm}\KK{missing}. The main advantage that \ourtool{} possesses for large systems is scalability in verifying large software systems through the abstraction processes mentioned above.

\begin{figure}[t]
  \includegraphics[width=0.9\linewidth]{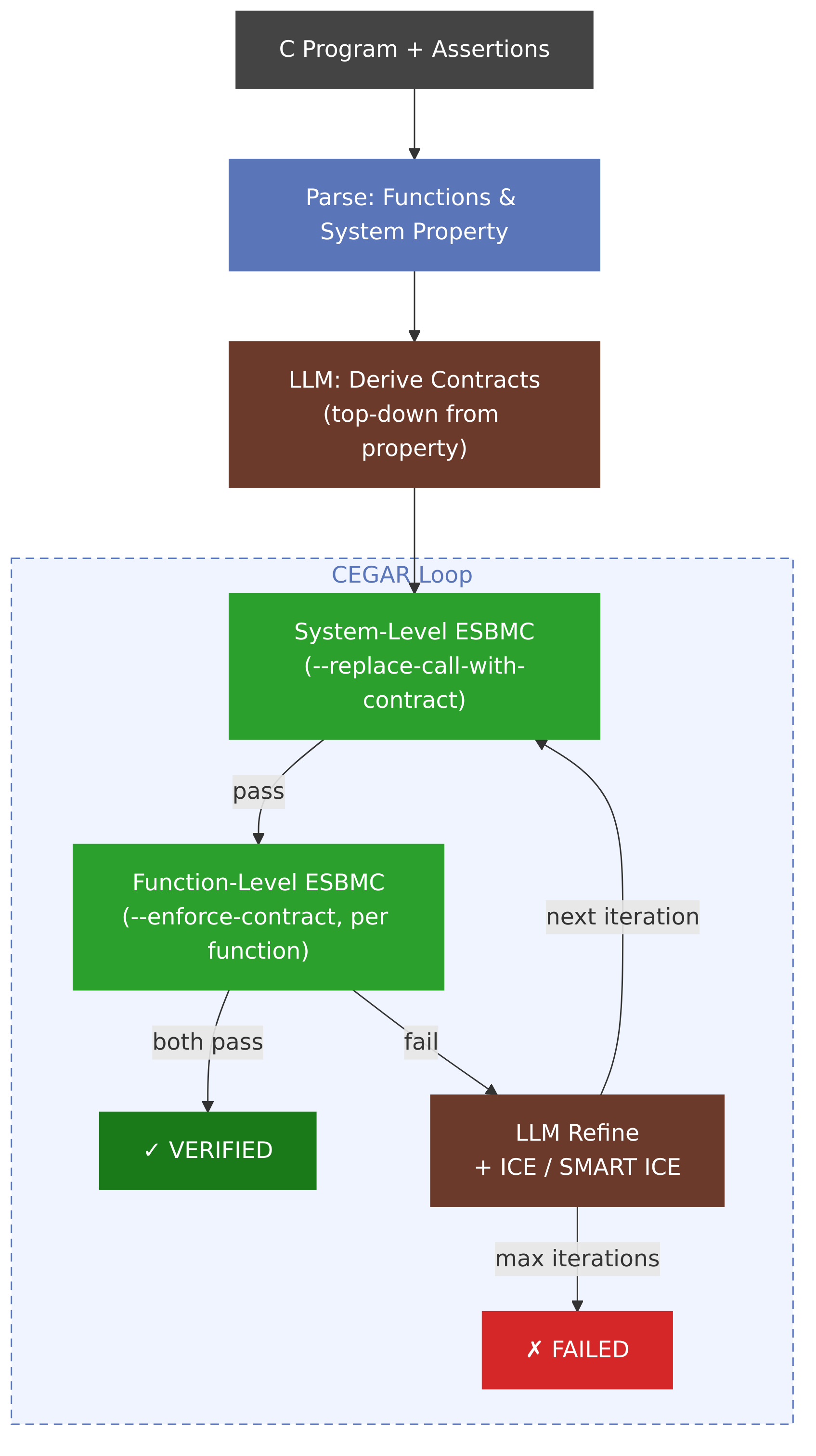}
  \caption{The \ourtool{} verification pipeline. The LLM derives function contracts top-down from the system property; the CEGAR loop alternates system-level and function-level ESBMC checks, refining contracts via ICE learning until both levels pass.}
  \label{fig:architecture}
\end{figure}

\paragraph{Contributions.} The main contributions of this paper are:
\begin{enumerate}
    \item \textbf{Top-down contract derivation.} A methodology that derives function contracts directly from a system-level assertion, minimising manual specification effort.
    \item \textbf{CEGAR with SMART ICE learning.} An iterative refinement loop that classifies ESBMC counterexamples into positive, negative, and implication examples to guide targeted LLM re-synthesis.
    \item \textbf{Loop invariant synthesis.} An on-demand inductive invariant generation step for functions where bounded checking is insufficient, integrated seamlessly into the Counterexample Guided Inductive Synthesis (CEGIS) loop.
    \item \textbf{End-to-end evaluation.} An empirical study across four benchmarks: the Frama-C benchmark (45~programs, 82--96\% success), LF2C-Simple (17~programs, 82--88\%), X.509 parser programs (6~programs, 33--50\%), and VerifyThis recursive/loop-intensive programs (11~programs, 55--64\%), covering diverse program structures and verification challenges.
    \item \textbf{\texttt{ESBMC-LF}.} A prototype preprocessor transpilation tool used to convert Lingua Franca (LF) benchmarks to C-based models, for verification with tools that work on the C programming language. Created for LF-Hard, comprising 24 LF Benchmarks that are converted into a monolithic C version. With the operational models of the LF Reactor replacing the implementation, this therefore simplifies the SMT equation and focuses on encoding only the model to be verified. Making the verification process simpler. Used as a Preprocessor tool to enable \ourtool{} to verify LF benchmarks.
    %could we change to "The operational model replaces the LF runtime, focusing verification on reaction logic only"
\end{enumerate}

We investigate three research questions throughout the paper.

\begin{dialogbox}[title=\textbf{RQ1}]
    \textit{How effectively does \ourtool{} verify programs across diverse benchmarks, and how does performance compare across three LLM backends?}
\end{dialogbox}

\begin{dialogbox}[title=\textbf{RQ2}]
    \textit{Under what conditions does SMART ICE learning improve convergence, and how does the quality of counterexample feedback interact with LLM capability?}
\end{dialogbox}

\begin{dialogbox}[title=\textbf{RQ3}]
    \textit{How does \ourtool{} scale to larger, structurally complex programs?}
\end{dialogbox}

The remainder of the paper is organised as follows.
Section~\ref{sec:background} covers background and Section~\ref{sec:related_work} covers similar research to \ourtool{}.
Section~\ref{sec:methodology} presents the \ourtool{} methodology.
Section~\ref{sec:benchmarks} presents the benchmarks that \ourtool{} will be tested on, along with the \texttt{ESBMC-LF} preprocessor.
Section~\ref{sec:experiments} reports experimental results.
Section~\ref{sec:discussion} discusses findings and threats to validity.
Section~\ref{sec:conclusion} discusses conclusions from the research.

%%%%%%%%%%%%%%%%%%%%%%%%%%%%%%%%%%%%%%%%%%%%%%%%%%%%%%%%%%%%%
\section{Background}
\label{sec:background}
%%%%%%%%%%%%%%%%%%%%%%%%%%%%%%%%%%%%%%%%%%%%%%%%%%%%%%%%%%%%

%As software systems continue to grow in size and complexity, the challenge of ensuring their correctness becomes increasingly critical. Program verification tools based on Bounded Model Checking (BMC), such as CBMC~\cite{cbmc-s2n} and ESBMC~\cite{wu2025esbmc}, provide powerful automated reasoning, but are limited by scalability. Function contracts offer a principled route to modular, scalable verification: they allow large verification tasks to be decomposed into smaller, independently checkable units. Projects such as AWS \texttt{s2n-tls}~\cite{cbmc-s2n} demonstrate the value of contracts in practice, yet also highlight that writing them manually demands significant expert effort. \ourtool{} automates this process by combining LLM-driven contract synthesis with formal CEGAR-based refinement. Additionally benefiting from the automated aspect of \ourtool{} is a lower barrier of entry for any prospective user. 

\paragraph{Bounded Model Checking (BMC)}~\cite{biere1999symbolic, biere2009bounded} verifies program correctness by unrolling loops up to a bound $k$ and encoding the resulting acyclic program as an SMT formula: a satisfying assignment corresponds to a counterexample, while unsatisfiability proves correctness up to bound $k$. ESBMC~\cite{cordeiro2010bounded, gadelha2018esbmc,  menezes2024esbmc, wu2025esbmc} is a mature BMC tool that supports C/C++ with rich arithmetic and pointer reasoning. The fundamental limitation of BMC is state space explosion: for large programs with many functions and deep call stacks, the SMT instance grows rapidly and solver time increases rapidly. Function contracts are the standard mitigation technique. Once a function is verified against its contract, the contract replaces the implementation in all calling contexts, preventing the solver from re-exploring the function's internals.

\paragraph{Large Language Models.} Recent work demonstrates that LLMs can reason effectively about code structure and semantics~\cite{jelodar2026large, xia_automated_2023, liu_evaluating_2024, tihanyi2025new, zubair_use_2025, haque_towards_2025}. LLMs have been applied to test generation, bug finding, and program synthesis, and more recently to formal specification tasks such as precondition and postcondition derivation~\cite{https://doi.org/10.34727/2025/isbn.978-3-85448-084-6_10, wang2025tale} \ourtool{} uses LLMs solely for contract and invariant synthesis for proposing candidate formal annotations, but delegates verification decisions to ESBMC, making the verification sound. This separation ensures that soundness depends only on the model checker, not on the LLM's output.

%%%%%%%%%%%%%%%%%%%%%%%%%%%%%%%%%%%%%%%%%%%%%%%%%%%%%%%%%%%%%
\section{Related Work}
\label{sec:related_work}
%%%%%%%%%%%%%%%%%%%%%%%%%%%%%%%%%%%%%%%%%%%%%%%%%%%%%%%%%%%%

\paragraph{PolyVer}
Many real-world software systems are \emph{polyglot}: their components are implemented in different programming languages, yet traditional verifiers target a single language, making whole-system verification challenging. PolyVer~\cite{https://doi.org/10.34727/2025/isbn.978-3-85448-084-6_10} addresses this by combining abstraction, compositional reasoning, and contract synthesis to verify polyglot systems. The system is modelled as a transition system whose components may be coded in different languages (e.g., C or Rust); PolyVer connects a top-level model checker (UCLID5~\cite{seshia2018uclid5}) with language-specific back-ends (CBMC for C, Kani for Rust) via automatically synthesised precondition/postcondition contracts at each language boundary. \ourtool{} is thematically related: both tools decompose cross-component verification through contracts, but \ourtool{} operates within a single language (C) and automates contract synthesis via LLMs rather than requiring manual annotation.
Additionally, \ourtool{} uses \texttt{ESBMC-LF} to create C-based models of LF benchmarks to automatically verify them.

\paragraph{ACSE}
Array-Carrying Symbolic Execution (ACSE)~\cite{lu2026array} automates function contract generation for array-manipulating C programs via a symbolic execution framework that carries invariants and contiguous array segments along execution paths. ACSE propagates loop invariants from external generators, handles disjunctive loop exits, and merges or splits array segments to synthesise precise ACSL pre/postconditions and assigns clauses, which are then discharged by Frama-C's WP plugin. \ourtool{} is complementary: both tools target automated contract synthesis, but ACSE derives contracts analytically in a single symbolic pass and crucially supports quantified array invariants (e.g., $\forall i.\; a[i] \geq 0$) that ESBMC's contract mechanism cannot express. \ourtool{} instead synthesises contracts via LLMs and iteratively refines them against formal counterexamples in a CEGAR/CEGIS loop, handling cases where symbolic invariant inference is imprecise or the program is too large for exhaustive symbolic exploration.

%%%%%%%%%%%%%%%%%%%%%%%%%%%%%%%%%%%%%%%%%%%%%%%%%%%%%%%%%%%%%
\section{Methodology}
\label{sec:methodology}
%%%%%%%%%%%%%%%%%%%%%%%%%%%%%%%%%%%%%%%%%%%%%%%%%%%%%%%%%%%%%

This Section will describe the functionality behind \ourtool{} and provide an overview of the function contract and loop invariant integration. Algorithm~\ref{alg:conver} covers how \ourtool{} works. \ourtool{} verifies a C program compositionally by decomposing its top-level system property into per-function contracts that are independently checkable by a formal verification tool. This abstraction of the state space enables verification of programs larger than would be possible with traditional formal verifiers due to state explosion. The verification happens at the following levels:

\begin{itemize}
    \item At the \emph{system level} \textit{(Step 3)}, every function call is replaced by a contract: ESBMC checks whether the system assertion holds assuming each function satisfies its postcondition, without exploring any function internals.
    \item At the \emph{function level} \textit{(Step 4)}, each implementation is checked against its own contract in isolation: ESBMC assumes the precondition holds and asserts the postcondition after execution, verifying the actual source code of that function.
\end{itemize}

%%%%%%%%%%%%%%%%%%%%%%%%%%%%%%%%%%%%%%%%%%%%%%%%%%%%%%%
\begin{algorithm}[h]
\caption{Overview of \ourtool{}}
\label{alg:conver}
\begin{algorithmic}[1]
\REQUIRE Program $P$, complexity threshold $\tau$, max iterations $K$
\ENSURE Verified $\top$ or Falsified $\bot$
%%%%%%%%%%%%%%%%%%%%%%%%%%%%%%%%%%%%%%%%%%%%%%%%%%%%%%%

\STATE \textbf{Step 1 --- Derive Sets} \label{alg:conver:step_1}
\STATE $\varphi \leftarrow \textsc{ExtractProperty}(P)$
\STATE $\mathcal{F} \leftarrow \textsc{ExtractFunctions}(P) \setminus \{\textsc{main}\}$
\STATE $\mathcal{F}_{\mathcal{L}} = \{f \in \mathcal{F} \mid \sigma(f) < \tau\}$ \CMT{Low complexity functions}
$\mathcal{F}_{\mathcal{H}} = \{f \in \mathcal{F} \mid \sigma(f) \geq \tau\}$ \CMT{High complexity functions}

\STATE \textbf{Step 2 --- Preprocessing} \label{alg:conver:step_2} \CMT{Run if Pre-Abstraction is enabled}
\STATE $\mathcal{C}_\mathcal{L} \leftarrow \textsc{Synthesize}(\mathcal{F}_{\mathcal{L}}, \varphi)$ \CMT{Simple; can verify directly}
\STATE $\mathcal{C}_\mathcal{H} \leftarrow \textsc{Overapproximate}(\mathcal{F}_{\mathcal{H}}, \varphi)$ \CMT{State explosion; too complex to verify directly}
\STATE $\mathcal{C} \leftarrow \mathcal{C}_\mathcal{L} \cup \mathcal{C}_\mathcal{H}$

\STATE \textbf{Step 3 --- System-Level Verification}\label{alg:conver:step_3}
\STATE $\mathcal{V}_S \leftarrow \textsc{VerifySystem}(P, \mathcal{C})$

\STATE \textbf{Step 4 --- Function-Level Verification} \label{alg:conver:step_4}
\STATE $\mathcal{V}_f \leftarrow \textsc{VerifyFunction}(P, \mathcal{C}, f)$ \CMT{For all {$f \in \mathcal{F}$}}

\IF{$\textsc{Pass}(\mathcal{V}_S) \land \forall f \in \mathcal{F}: \textsc{Pass}(\mathcal{V}_f)$}
    \RETURN $\top$ \CMT{Contracts correct without refinement}
\ENDIF

\STATE \textbf{Step 5 --- Contract Replacement and System Verification} \label{alg:conver:step_5}
\STATE $\mathcal{C} \leftarrow \{C_f \in \mathcal{C} \mid \textsc{Pass}(\mathcal{V}_f)\}$ \CMT{Drop contracts that failed function-level verification}
\STATE $\mathcal{V}_S \leftarrow \textsc{VerifySystem}(P, \mathcal{C})$ \CMT{Verify using surviving contracts}

\STATE \textbf{Step 6 --- CEGAR Refinement} \label{alg:conver:step_6}
\STATE $\langle \mathcal{C}, \mathcal{V}_S, \mathcal{V}_F \rangle \leftarrow \textsc{CEGAR}(\mathcal{C}, \mathcal{V}_S, \mathcal{V}_F, K)$ \CMT{See Alg.~\ref{alg:cegar}}
\IF{$\textsc{Pass}(\mathcal{V}_S) \land \forall v \in \mathcal{V}_F: \textsc{Pass}(v)$}
    \RETURN $\top$
\ENDIF

\STATE \textbf{Step 7 --- CEGIS Escalation} \label{alg:conver:step_7}
\STATE $\langle \mathcal{C}, \mathcal{V}_S, \mathcal{V}_F \rangle \leftarrow \textsc{CEGIS}(\mathcal{C}, \mathcal{V}_S, \mathcal{V}_F, K)$ 
\CMT{See Alg.~\ref{alg:cegis}}
\IF{$\textsc{Pass}(\mathcal{V}_S) \land \forall v \in \mathcal{V}_F: \textsc{Pass}(v)$}
    \RETURN $\top$
\ENDIF

\RETURN $\bot$

\end{algorithmic}
\end{algorithm}
%%%%%%%%%%%%%%%%%%%%%%%%%%%%%%%%%%%%%%%%%%%%%%%%%%%%%%%

ConVer uses a two-stage refinement strategy. Both stages verify contracts at the system level (proving the property holds assuming contracts --- Step 3) and the function level (proving each implementation satisfies its contract --- Step 4). The CEGAR loop first attempts to refine contracts using LLM-guided counterexample analysis (Step 5). If CEGAR stagnates, a variant of Delta Debugging~\cite{zeller1999yesterday, zeller2002simplifying} is used as a stagnation breaker by incrementally removing post-condition clauses from stagnating functions, reverifies said function until the verification passes, then pipeline escalates to CEGIS (Step 7). CEGIS performs constraint-based synthesis against accumulated positive and negative examples. LLMs are integrated into both loops to facilitate contract generation.

 % \begin{dialogbox}[title=Compositional Soundness]
 %     When both levels pass, the disjunction of the verification conditions is sound by composition: the original code satisfies the contracts (function-level), and the system property holds under those contracts (system-level), the system property holds for the original code. This argument holds regardless of how loose or tight the contracts are, provided the system-level check passes with the same contracts that the function-level check verifies. See Section~\ref{sec:func_contract} for an explanation.
 % \end{dialogbox}

%%%%%%%%%%%%%%%%%%%%%%%%%%%%%%%%%%%%%%%%%%%%%%%%%%%%%%%%%%%%%
\subsection{ESBMC Function Contract and Loop Invariant Checker}
\label{sec:func_contract}
%%%%%%%%%%%%%%%%%%%%%%%%%%%%%%%%%%%%%%%%%%%%%%%%%%%%%%%%%%%%%

ESBMC~\cite{wu2025esbmc} supports modular verification through two complementary mechanisms: \emph{function contracts} and \emph{loop invariants}. Function contracts support \emph{deductive} reasoning ---deriving postconditions from preconditions without re-examining the function body---while loop invariants support \emph{inductive} reasoning, establishing correctness for unbounded loops through a fixed base case and preservation argument. Together, they form the verification backbone of \ourtool{}, enabling large-scale verification tasks to be decomposed into independently checkable units.

\paragraph{Function Contracts.}
A function contract for a function $f$ is a triple $(P, Q, S)$, where 
$P : \mathit{State} \to \mathbb{B}$ is the precondition, 
$Q : \mathit{State} \times \mathit{State} \to \mathbb{B}$ is the postcondition 
relating pre- and post-states, and $S \subseteq \mathit{Vars}$ is the assigns 
set of locations $f$ is permitted to modify. In ESBMC these are expressed via 
\texttt{\_\_ESBMC\_requires}, \texttt{\_\_ESBMC\_ensures}, and 
\texttt{\_\_ESBMC\_assigns}. A simple example is shown below:
\begin{lstlisting}[language=C]
int increment(int x) {
    __ESBMC_requires(x > 0);
    __ESBMC_assigns(x);
    __ESBMC_ensures(__ESBMC_return_value > x);
    return x + 1;
}
\end{lstlisting}
Given that $P$ holds at a call site, ESBMC deduces that $Q$ holds upon return 
without re-examining the body; verification proceeds as 
$\texttt{assume}(P) \to \texttt{execute}(f) \to \texttt{assert}(Q)$. 
Once $f$ is verified to satisfy $(P, Q, S)$, the contract replaces the 
implementation in any calling context.

Function contracts serve two distinct roles in \ourtool{}. In 
\emph{system-level verification} (\emph{Replace} mode), each call to $f$ is 
substituted with an abstract stub
\begin{equation}
  \mathit{abs}(P,Q,s_0) \;=\; \{\,s \mid P(s_0) \wedge Q(s_0,s)\,\},
  \label{eq:abs}
\end{equation}
which havocs only the declared assigns targets and assumes $Q$; the internal state space is never explored. In \emph{function-level verification} (\emph{Enforce} mode), the real implementation is checked against $(P,Q,S)$ by assuming $P$ and asserting $Q$ after execution. Enforce mode establishes the \emph{contract-conformance} condition
\begin{equation}
  \forall s_0.\; P(s_0) \;\Rightarrow\; \llbracket f \rrbracket(s_0) 
  \;\in\; \mathit{abs}(P,Q,s_0),
  \label{eq:conformance}
\end{equation}
i.e., the concrete execution trace of $f$ is a witness in the abstract state set defined by the stub. Expanding~\eqref{eq:abs} into~\eqref{eq:conformance} reduces to the standard Hoare triple $\{P\}\,f\,\{Q\}$, which is precisely what ESBMC discharges during Enforce mode. A function that passes this check is \emph{contract-conformant}. When both checks pass, soundness follows by substitution: because the real implementation satisfies \eqref{eq:conformance}, any property $\phi$ proved against the stub in Replace mode holds for the real code as well, and the composed result is end-to-end sound.

\paragraph{Loop Invariants.}
For programs containing loops, ESBMC supports the annotation of \emph{loop invariants} via \\\texttt{\_\_ESBMC\_loop\_invariant()}~\cite{esbmc74}. Unlike function contracts, which rely on deduction, loop invariants establish correctness through \emph{induction}: a predicate is shown to hold before the loop begins, to be preserved by each iteration, and to imply the desired postcondition upon termination. Consider the following example:

\begin{lstlisting}[language=C]
int sum = 0;
for (int i = 0; i < n; i++) {
    __ESBMC_loop_invariant(sum == i * (i - 1) / 2);
    sum += i;
}
\end{lstlisting}

The invariant $\texttt{sum} = \frac{i(i-1)}{2}$ is verified inductively by ESBMC in three steps:
\begin{enumerate}
    \item \emph{Base case}: the invariant holds at $i = 0$, since $\texttt{sum} = 0 = \frac{0 \cdot (-1)}{2}$.
    \item \emph{Inductive step}: assuming the invariant holds at iteration $i$, after executing \texttt{sum += i}, we obtain $\texttt{sum}' = \frac{i(i-1)}{2} + i = \frac{i(i+1)}{2}$, which matches the invariant at $i' = i+1$.
    \item \emph{Conclusion}: when the loop exits at $i = n$, the invariant yields $\texttt{sum} = \frac{n(n-1)}{2}$.
\end{enumerate}

\begin{dialogbox}
    This reduces an otherwise unbounded verification problem to a fixed inductive argument, avoiding loop unrolling entirely and sidestepping the state space explosion that would otherwise arise for large or non-terminating loops \cite{pirzada2024llm}.
\end{dialogbox}

\subsection{CEGAR+CEGIS}
%%%%%%%%%%%%%%%%%%%%%%%%%%%%%%%%%%%%%%%%%%%%%%%%%%%%%%%%%%%%%

\paragraph{Counterexample-Guided Abstraction Refinement (CEGAR)}
\cite{clarke2000counterexample} drives the primary refinement loop in \ourtool{}. Starting from LLM-derived contracts, each iteration runs two ESBMC checks: System-Level verification (function calls replaced by abstract contract stubs) and Function-Level verification (each implementation checked against its contract). If both pass, verification is sound by the assume-guarantee rule~\cite{komuravelli2012assume}. Otherwise, the counterexample is fed back to the LLM for contract refinement. Function-level failures direct the LLM to relax contracts; system-level failures direct it to strengthen them --- reflecting that the two levels pull contracts in opposite directions. CEGAR runs for up to five iterations before escalating. An overview of CEGAR used in \ourtool{} can be seen in Algorithm~\ref{alg:cegar}.

\begin{algorithm}[ht]
\caption{CEGAR Refinement in \ourtool{}} 
\label{alg:cegar}
\begin{algorithmic}[1] 
\REQUIRE Contracts $\mathcal{C}$, results $\mathcal{V}_S, \mathcal{V}_F$, max iterations $K$
\ENSURE Refined $\langle \mathcal{C}, \mathcal{V}_S, \mathcal{V}_F \rangle$ 
\FOR{$k = 1$ \TO $K$}
    \STATE $\mathcal{C} \leftarrow \textsc{CounterexampleRefine}(\mathcal{C}, \mathcal{V}_S, \mathcal{V}_F)$ 
    \STATE $\mathcal{V}_S \leftarrow \textsc{VerifySystem}(P, \mathcal{C})$ 
    \STATE $\mathcal{V}_f \leftarrow \textsc{VerifyFunction}(P, \mathcal{C}, f)$ for all $f \in \mathcal{F}$ 
    \IF{$\textsc{Pass}(\mathcal{V}_S) \land \forall v \in \mathcal{V}_F: \textsc{Pass}(v)$} 
        \RETURN $\langle \mathcal{C}, \mathcal{V}_S, \mathcal{V}_F \rangle$ 
    \ENDIF 
    \IF{$\textsc{Stagnant}(k)$} 
        \STATE $\mathcal{C} \leftarrow \textsc{DeltaDebug}(\mathcal{C})$ 
        \STATE \textbf{break} \CMT{Escalate to CEGIS --- Step 7}
    \ENDIF
\ENDFOR 
\RETURN $\langle \mathcal{C}, \mathcal{V}_S, \mathcal{V}_F \rangle$
\end{algorithmic}
\end{algorithm} 

\paragraph{Counterexample-Guided Inductive Synthesis (CEGIS)}
\cite{10.1145/1375581.1375599} serves as a second-stage fallback when CEGAR exhausts its budget. Counterexamples are classified and accumulated in a constraint database tracking positive examples (passing configurations), negative examples (states that must not satisfy the contract), and implication examples (inferred from observed program behaviour and 
structural patterns in the function body). Each synthesis call is conditioned on the full, accumulated constraint set, guiding the LLM toward contracts that are consistent with all observed behaviour. This loop also runs for up to five iterations; if the combined budget of up to ten iterations across both stages is exhausted without convergence, verification is declared inconclusive. Algorithm~\ref{alg:cegis} describes how \ourtool{} uses CEGIS.

\begin{algorithm}[ht]
\caption{CEGIS Escalation in \ourtool{}}
\label{alg:cegis} 
\begin{algorithmic}[1]
\REQUIRE Contracts $\mathcal{C}$, results $\mathcal{V}_S, \mathcal{V}_F$, max iterations $K$ 
\ENSURE Refined $\langle \mathcal{C}, \mathcal{V}_S, \mathcal{V}_F \rangle$ 
 
\STATE $\mathcal{D} \leftarrow \textsc{CEGISDatabase}(\mathcal{V}_F)$ \CMT{Migrate counterexamples from CEGAR} 
\FOR{$k = 1$ \TO $K$}
    \STATE $\mathcal{C} \leftarrow \textsc{ExampleSynthesize}(\mathcal{C}, \mathcal{D})$ \CMT{Using Oracle} 
    \STATE $\mathcal{V}_S \leftarrow \textsc{VerifySystem}(P, \mathcal{C})$ 
    \STATE $\mathcal{V}_f \leftarrow \textsc{VerifyFunction}(P, \mathcal{C}, f)$ for all $f \in \mathcal{F}$ 
    \IF{$\textsc{Pass}(\mathcal{V}_S) \land \forall v \in \mathcal{V}_F: \textsc{Pass}(v)$} 
        \RETURN $\langle \mathcal{C}, \mathcal{V}_S, \mathcal{V}_F \rangle$ 
    \ENDIF 
    \STATE Update $\mathcal{D}$ with new $E^+, E^-$ from $\mathcal{V}_F$ 
\ENDFOR 
\RETURN $\langle \mathcal{C}, \mathcal{V}_S, \mathcal{V}_F \rangle$ 
\end{algorithmic} 
\end{algorithm} 

\subsection{SMART ICE}
%%%%%%%%%%%%%%%%%%%%%%%%%%%%%%%%%%%%%%%%%%%%%%%%%%%%%%%%%%%%%

\paragraph{ICE Learning.}
\ourtool{} structures verification feedback using the ICE framework~\cite{garg2014ice}, which classifies examples into three categories: positive examples~($E^{+}$) of states that must be included in the contract, negative examples~($E^{-}$) of states that must be excluded, and implication examples encoding inductive relationships between states. Conflict detection identifies when a counterexample contradicts existing positive examples, keeping the database consistent. When ESBMC reports a failure, the raw output is parsed to extract the violated property and execution trace; tool-level failures are separated from semantic counterexamples, which are further classified into five categories --- only unconstrained and semantic violations are admitted into~$E^{-}$, keeping the database free of noise.

\begin{algorithm}
\caption{Counterexample Handling Pipeline - SMART ICE}
\begin{algorithmic}[1]
\REQUIRE Raw ESBMC output, current contracts, ICE database
\ENSURE Updated $E^{+}$/$E^{-}$/implications, weakest-link function
\STATE Parse output: extract violated property, trace, key variables
\IF{tool-level failure (timeout / internal error / parse rejection)}
    \STATE Record separately; \textbf{exit}
\ENDIF
\STATE Classify into: \textit{syntax error}, \textit{unparsed}, \textit{tool error}, \textit{unconstrained init}, \textit{semantic}
\IF{semantic or unconstrained initialisation}
    \STATE Admit into $E^{-}$
\ENDIF
\IF{system-level failure}
    \STATE Map variables $\rightarrow$ responsible functions
    \STATE Compute gap score per function
    \STATE Flag the highest-gap function as the weakest link
\ENDIF
\STATE Render structured diagnostic block (trace $+$ $E^{+}$/$E^{-}$ analysis) for LLM prompt
\end{algorithmic}
\end{algorithm}

\subsection{Pre-abstraction}
%%%%%%%%%%%%%%%%%%%%%%%%%%%%%%%%%%%%%%%%%%%%%%%%%%%%%%%%%%%%%

The pre-abstraction phase addresses the scalability limitation of compositional verification: certain functions exhibit structural complexity sufficient to trigger state explosion even at the function level. \ourtool{} assigns each domain function a \textit{complexity score} which is a weighted sum of static metrics extracted by regex analysis of the function body following established complexity metrics~\cite{mccabe1976complexity}. The weights reflect the cost model of BMC: loop count and nesting depth carry the highest per-occurrence weights because each iteration multiplies the path count. Recursion and unbounded loops incur large flat penalties, as a single occurrence renders the state space potentially infinite. Dynamic allocation is penalised moderately (bounded by calling context); branching and pointer operations contribute smaller per-occurrence costs.

The score maps each function to one of four explosion-risk tiers:
\textit{minimal} ($<5.0$), \textit{low} ($\geq 5.0$), \textit{medium}
($\geq 10.0$), and \textit{high} ($\geq 20.0$). Functions in the \textit{medium} or \textit{high} tier (score $\geq 10.0$, configurable) are designated for \textit{over-approximate abstraction}; the rest undergo standard precise-contract synthesis. The LLM prompt for abstraction instructs the model to produce
\textit{sound but loose} contracts: weaker postconditions (null-safety, range bounds) rather than full behavioural specifications, and an over-approximate \texttt{assigns} clause covering all globals that \emph{might} be modified. A heuristic fallback generates tautological \texttt{ensures} clauses and a body-scan-derived \texttt{assigns} list when the LLM fails or produces unparseable output.

The pipeline then proceeds through five phases:
\begin{itemize}
    \item \textbf{Phase~1a}: Generate over-approximate abstractions for
          high-complexity functions.
    \item \textbf{Phase~1b}: Concurrently synthesise precise contracts for
          low-complexity functions via coverage-validated LLM derivation.
    \item \textbf{Phase~2}: Run an initial ESBMC system-level check using the
          abstractions as function summaries, yielding a fast over-approximate
          verification result.
    \item \textbf{Phase~3}: Verify each function atomically against its own
          contract (proceeds unconditionally, regardless of Phase~2 outcome).
    \item \textbf{Phase~4}: Replace each successfully verified abstraction with
          its precise contract; unverified functions retain their conservative
          abstraction.
    \item \textbf{Phase~5}: Escalate to CEGAR/CEGIS refinement if the
          substituted contracts still fail to establish the system property.
\end{itemize}

This stratified strategy --- abstract first, verify atomically, then refine ---
allows \textsc{ConVer} to handle programs that would otherwise be intractable
under a single-phase compositional approach.

\section{Benchmarks}
\label{sec:benchmarks}
%%%%%%%%%%%%%%%%%%%%%%%%%%%%%%%%%%%%%%%%%%%%%%%%%%%%%%%%%%%%%

This Section describes the benchmarks that \ourtool{} verifies.

%%%%%%%%%%%%%%%%%%%%%%%%%%%%%%%%%%%%%%%%%%%%%%%%%%%%%%%%%%%%
\subsection{\texttt{ESBMC-LF} and LF-Hard}
\label{subsec:lf2cbench}
%%%%%%%%%%%%%%%%%%%%%%%%%%%%%%%%%%%%%%%%%%%%%%%%%%%%%%%%%%%%%

To directly compare ConVer with tools that consume LF files, such as work from~\cite{https://doi.org/10.34727/2025/isbn.978-3-85448-084-6_10} which uses the LF Verfifier Benchmarks~\cite{lin2023towards}, a purpose-built preprocessor tool \texttt{ESBMC-LF} was created to convert the LF file benchmarks to C. \texttt{ESBMC-LF} translates the LF files into C with the verification conditions included, described in Algorithm~\ref{alg:lf2c}. The purpose of the pipeline is to preserve the verification constructs of the LF benchmarks, which the LFC compiler strips from the converted code to ensure the compiled program runs efficiently. The resulting LF-Hard suite comprises 24~C programs converted from the LF Verifier Benchmarks~\cite{lin2023towards}; further discussed in Section~\ref{subsec:benchmarks}. The remainder of Section~\ref{subsec:lf2cbench} describes in further detail Algorithm~\ref{alg:lf2c}.

\textbf{Steps~1--3} compile each LF benchmark to C via the LFC transpiler, normalize platform-dependent paths and compiler options, and combine all source files into a single translation unit. During combination, the reactor-c runtime library is replaced with a verification-friendly operational model that stubs the scheduler, event queue, and token lifecycle, so that ESBMC reasons only about the benchmark's reaction logic.

\textbf{Steps~4--5} extract the Metric Temporal Logic (MTL) properties from the original LF source and encode them as ESBMC verification constructs (\texttt{\_\_ESBMC\_assert}, \texttt{\_\_ESBMC\_assume}, and tracking variables) in the combined C file. Each MTL formula is pattern-matched to one of nine instrumentation templates that determine where assertions are placed and whether tracking variables are needed.
% See Appendix~ref{sec:app:lf2c} for a detailed description of these transformations.

\textbf{Steps~6--7} generate a direct execution model that replaces the reactor-c scheduler. The pipeline parses the output topology to build a cascade graph; a static map from each reaction's output port to its downstream reactions and emits direct function calls with \texttt{is\_present} guards instead of runtime dispatch through trigger arrays. Loop bounds are derived from timer periods and property time horizons, yielding a finite, fully unrollable execution loop. Dead runtime infrastructure is then stripped to reduce the symbolic state space.

%%%%%%%%%%%%%%%%%%%%%%%%%%%%%%%%%%%%%%%%%%%%%%%%%%%%%%%%%%%%%
\begin{algorithm}
\caption{\texttt{ESBMC-LF}: Converts LF files to C with verification property preservation}
\label{alg:lf2c}
%%%%%%%%%%%%%%%%%%%%%%%%%%%%%%%%%%%%%%%%%%%%%%%%%%%%%%%%%%%%%
\begin{algorithmic}[1]
    \REQUIRE LF Benchmark $B$
    \ENSURE C Program $\mathit{c\_file}$ with property set $\Phi$

    \STATE \textbf{Step 1: Semantic LF-to-C translation}
    \STATE $P \gets \textsc{CompileLF}(B)$
        \CMT{$P$ is a C project semantically equivalent to $B$ without MTL properties assuming \texttt{LFC} tool performs sound conversion}

    \STATE \textbf{Step 2: Platform normalization}
    \STATE $P' \gets \textsc{NormalizeProject}(P)$
        \CMT{Resolve platform-dependent paths, headers, and definitions}

    \STATE \textbf{Step 3: Single translation unit construction}
    \STATE $C \gets \textsc{FlattenProject}(P',\ \mathit{OM})$
        \CMT{Replace runtime library with operational model $\mathit{OM}$, preprocess into one translation unit, filter system headers}

    \STATE \textbf{Step 4: Property extraction}
    \STATE $\Phi \gets \textsc{ExtractMTL}(B,\ C)$
        \CMT{Parse MTL annotations from $B$, map symbols to locations in $C$}

    \STATE \textbf{Step 5: Property instrumentation}
    \STATE $C' \gets \textsc{InstrumentMTL}(\Phi,\ C)$
        \CMT{Insert assertions, assumptions, and tracking variables per $\Phi$}

    \STATE \textbf{Step 6: Execution model generation}
    \STATE $C'' \gets \textsc{GenExecModel}(C')$
        \CMT{Build cascade graph from output topology, replace scheduler dispatch with direct calls, derive loop bounds}

    \STATE \textbf{Step 7: Post-processing}
    \STATE $\mathit{c\_file} \gets \textsc{PostProcess}(C'')$
        \CMT{Strip dead runtime code, fix preprocessor-baked constants, propagate action values, remove line directives}

    \RETURN $\mathit{c\_file}$
\end{algorithmic}
\end{algorithm}
%%%%%%%%%%%%%%%%%%%%%%%%%%%%%%%%%%%%%%%%%%%%%%%%%%%%%%%%%%%%%

%%%%%%%%%%%%%%%%%%%%%%%%%%%%%%%%%%%%%%%%%%%%%%%%%%%%%%%%%%%%%
\subsection{Benchmark Suites}
\label{subsec:benchmarks}
%%%%%%%%%%%%%%%%%%%%%%%%%%%%%%%%%%%%%%%%%%%%%%%%%%%%%%%%%%%%%

We evaluate \ourtool{} on four alternative benchmark suites totalling 79~C programs, spanning algorithmic code, loop and recursion intensive verification challenges, and real-world parsing routines.

\paragraph{LF2C-Simple (17 Programs).}
These LF2C-Simple Benchmarks were synthesised using Claude 3.5 Sonnet \cite{anthropic2024claude35sonnet}. Where the LLM was prompted with the original LF Benchmark with the explicit instruction to generate a semantically equivalent C Program version with the system property inserted at the correct position inside the code. We successfully obtained 17 of the 24 benchmarks; the remainder contained buggy cases that led to Verification Failed due to a logic error injected by the LLM, and we were unable to verify the system property.

\paragraph{Frama-C (45 programs).}
Drawn from a Frama-C/ACSL exercise set used to evaluate deductive verification tools at FM\,2026-AE~\cite{lu2026array}, the suite covers numerical computations, array operations, sorting, and searching (e.g., \texttt{add}, \texttt{sort}, \texttt{binary\_search}). Original ACSL annotations (\texttt{//@ requires}/\texttt{ensures}) are removed, and the same property is re-expressed as a single top-level \texttt{assert()} in \texttt{main()}. Programs are typically under 100~LOC, making this suite well-suited for evaluating the core CEGAR plus ICE contract synthesis pipeline on well-structured algorithmic code without structural interference.

\paragraph{VerifyThis (11 programs).}
Drawn from three SV-COMP~\cite{beyer2024state} categories --- six from the VerifyThis track (\texttt{lcp}, \texttt{prefixsum}, \texttt{prefixsum\_rec}, \texttt{duplets}, \texttt{elimination\_max\_rec}, \texttt{elimination\_max\_rec\_onepoint}), two from recursive-simple (\texttt{fibo}, \texttt{sum}), and three from recursive (\texttt{MultCommutative}, \texttt{EvenOdd}, \texttt{Addition}) --- the programs are adapted by replacing dynamic memory allocation with fixed-size stack arrays and reducing verification bounds (e.g., $n \leq 8$ for prefix sum, $n \leq 6$ for elimination max) to prevent state-space explosion. Four programs were excluded: \texttt{elimination\_max} has a monolithic \texttt{main()} with no callee, and three tree-manipulation programs (\texttt{tree\_del\_iter}, \texttt{tree\_del\_rec}, \texttt{tree\_max}) require heap-allocated structures with $\forall$-quantified postconditions unsupported by \newline \texttt{\_\_ESBMC\_ensures}. The suite is evaluated exclusively with the Pre-Abstraction strategy, as loop-intensive programs cause state-space explosion under direct synthesis and recursive programs exceed SmartICE's capacity without prior abstraction.

\paragraph{X.509 (6 programs).}
Extracted from the ANSSI-FR \texttt{x509-parser} project and included in the same FM\,2026 artifact~\cite{lu2026array} as the Frama-C suite, these programs implement real-world ASN.1/X.509 routines: buffer comparison (\texttt{bufs\_differ}), IA5 string validation \newline(\texttt{check\_ia5\_string}), ASN.1 type parsers (\texttt{parse\_null}, \newline \texttt{parse\_algoid\_params\_none}), time-field encoding \newline(\texttt{time\_components\_to\_comparable\_u64}), and time field checking (\texttt{verify\_correct\_time\_use}). Three adaptations are applied for ESBMC compatibility: (1)~\texttt{static} qualifiers are removed (required for \texttt{--enforce-contract} to locate functions by name); (2)~ACSL assertions are replaced with C \texttt{assert()}; and (3)~universal quantifiers are encoded via a nondet index---e.g., \texttt{\textbackslash forall integer i; 0<=i<len ==> buf[i]<=0x7f} becomes:

\begin{lstlisting}[language=C, numbers=none, frame=none, basicstyle=\small\ttfamily]
u32 idx;
__ESBMC_assume(idx < len);
if (ret == 0) assert(buf[idx] <= 0x7f);
\end{lstlisting}

ESBMC's BMC engine explores all nondet values of \texttt{idx}, making this encoding semantically equivalent to the universal quantifier within the bounded path. Ranging from single-function leaf parsers to three-level call chains with pointer arithmetic, this suite is substantially harder than the Frama-C benchmark and probes whether \ourtool{}'s CEGAR+ICE loop can synthesise sound pointer contracts for low-level code.

\section{Experiments}
\label{sec:experiments}
%%%%%%%%%%%%%%%%%%%%%%%%%%%%%%%%%%%%%%%%%%%%%%%%%%%%%%%%%%%%%

\paragraph{Tool and hardware configuration.}
All experiments use ESBMC~8.1.0 as the backend verifier and perform at most 5 CEGAR iterations per program.
We evaluate three LLM backends: \textit{Qwen-Plus} (DashScope cloud API), \textit{Claude Haiku~4.5} (Anthropic cloud API), and \textit{GPT-OSS~120b} served locally via Ollama on a server with four NVIDIA RTX~A6000 GPUs (48\,GB each).
Table~\ref{tab:setup} summarises the parameters.
For the Frama-C benchmark, LF2C-Simple, and X.509 we set a per-program timeout of 600\,s and run three workers in parallel for cloud models (one for GPT-OSS~120b).
For VerifyThis, we extend the timeout to $900$\,s to accommodate deeper recursion and loop complexity.

\begin{table}[t]
  \centering
  \footnotesize
  \caption{Experimental setup.}
  \label{tab:setup}
  \begin{tabular}{@{}p{0.62\columnwidth}p{0.32\columnwidth}@{}}
    \toprule
    Parameter & Value \\
    \midrule
    ESBMC version      & 8.1.0 \\
    LLM backends       &  Qwen3.5-Plus; Haiku~4.5; GPT-OSS~120b \\
    Max iterations     & 5 \\
    Timeout (Frama-C / Simple / X.509) & 600\,s \\
    Timeout (VerifyThis)               & 900\,s \\
    Workers (cloud / local)          & 3\,/\,1 \\
    \bottomrule
  \end{tabular}
\end{table}

\paragraph{Strategies.}
We compare two verification strategies: \emph{SmartICE} applies the full CEGAR+SMART~ICE pipeline with structured counterexample feedback, while \emph{No-ICE} disables ICE learning and serves as an ablation baseline.
Both strategies are applicable to the Frama-C benchmark, LF2C-Simple, and X.509, which consist of short programs with explicit function-call structure.
For VerifyThis --- programs with deep or unbounded recursion that cause state explosion under direct synthesis --- we apply the \emph{Pre-Abstraction} strategy exclusively.

% \paragraph{Research questions.}
% We address the following research questions:
% \begin{itemize}
%   \item[\textbf{RQ1}] How effectively does \ourtool{} verify programs across diverse benchmarks, and how does performance compare across three LLM backends?
%   \item[\textbf{RQ2}] Under what conditions does SMART ICE learning improve convergence, and how does the quality of counterexample feedback interact with LLM capability?
%   \item[\textbf{RQ3}] How does \ourtool{} scale to larger, structurally complex programs?
% \end{itemize}

%%%%%%%%%%%%%%%%%%%%%%%%%%%%%%%%%%%%%%%%%%%%%%%%%%%%%%%%%%%%%
\subsection{Results on Function-Contract Benchmarks (RQ1)}
%%%%%%%%%%%%%%%%%%%%%%%%%%%%%%%%%%%%%%%%%%%%%%%%%%%%%%%%%%%%%

\begin{figure*}[t]
  \includegraphics[width=\linewidth]{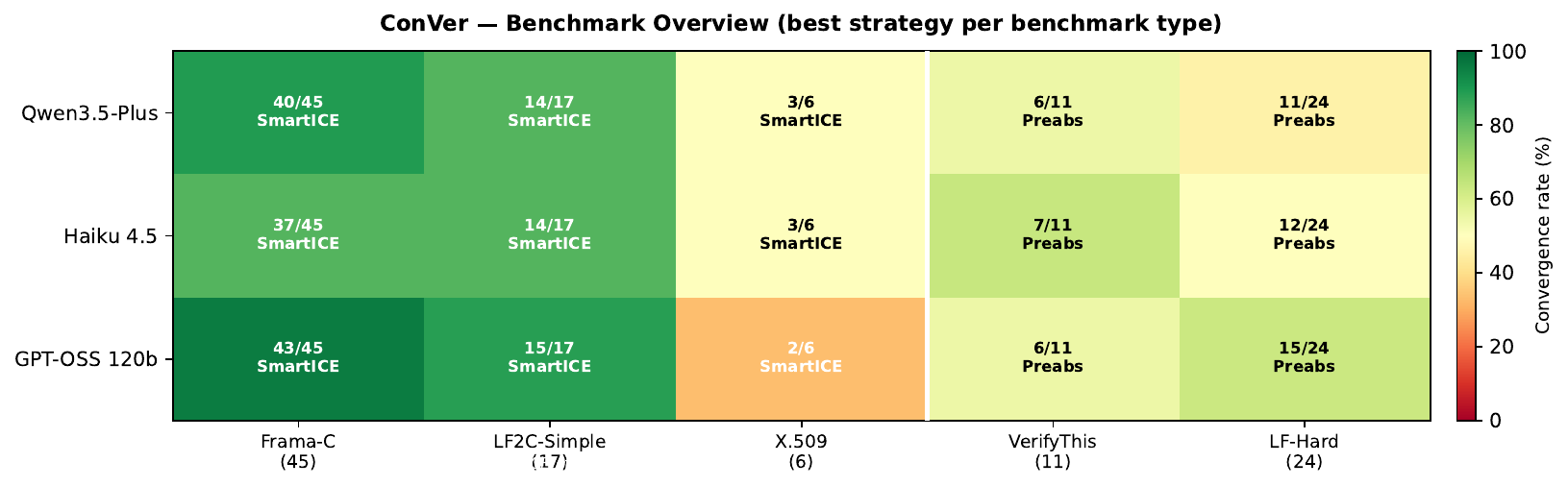}
  \caption{Verification success rate (\%) across all four benchmarks and three LLM backends.
    Each cell shows the convergence percentage under the best applicable strategy for that benchmark.}
  \label{fig:overview_heatmap}
\end{figure*}

Table~\ref{tab:main_results} shows verification outcomes on Frama-C benchmark (45 programs), LF2C-Simple (17 programs), and X509 (6 programs) using the SmartICE strategy.
Figure~\ref{fig:overview_heatmap} provides a visual overview across all four benchmarks.

\begin{table}[t]
  \centering
  \small
  \caption{Verified programs (SmartICE strategy) on the Frama-C benchmark, LF2C-Simple, and X509.
    Numbers show converged\,/\,total (\%).
    Qwen = Qwen-Plus; Claude = Haiku~4.5; GPT = GPT-OSS~120b.}
  \label{tab:main_results}
  \begin{tabular}{lccc}
    \toprule
    Benchmark        & Qwen   & Claude & GPT-OSS \\
    \midrule
    Frama-C (45)     & 40 (89\%)  & 37 (82\%)  & \textbf{43 (96\%)} \\
    LF2C-Simple (17) & 14 (82\%)  & 14 (82\%)  & \textbf{15 (88\%)} \\
    X509    (6)      & 3 (50\%)   & 3 (50\%)   & 2 (33\%) \\
    \bottomrule
  \end{tabular}
\end{table}

\begin{figure}[t]
  \includegraphics[width=\linewidth]{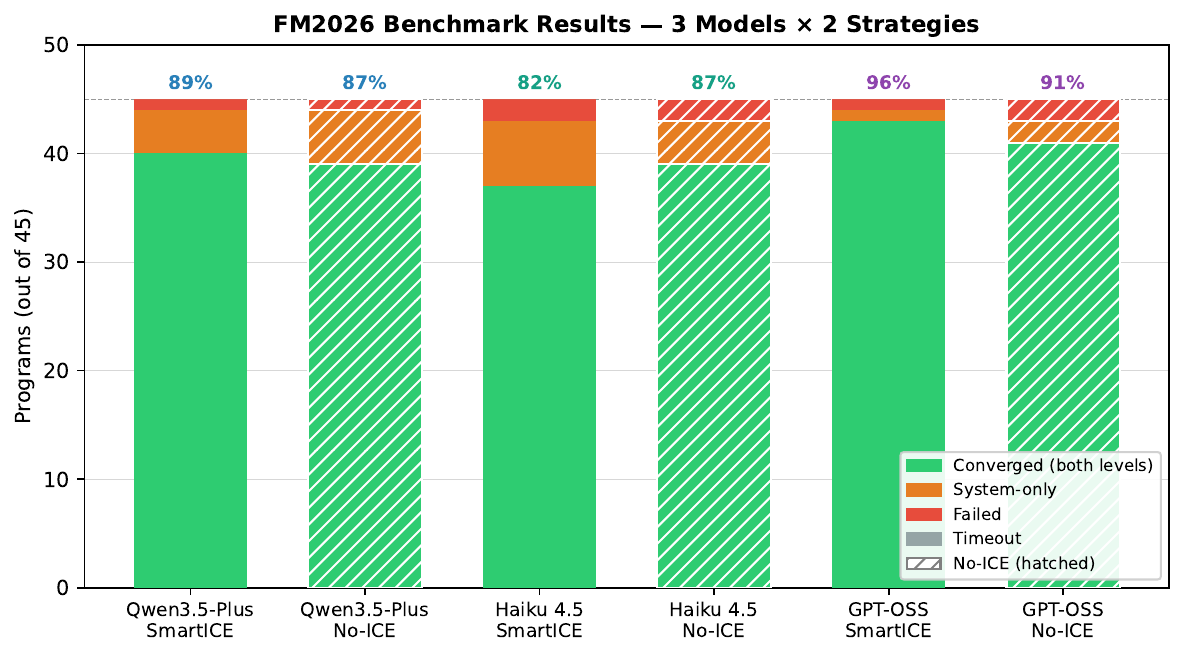}
  \caption{Outcome breakdown on Frama-C benchmark (45 programs) for three LLM backends and two strategies.
    Stacked bars show counts for converged (green), system-only (orange), failed (red), and timeout (grey).}
  \label{fig:fm2026_bar}
\end{figure}

\ourtool{} achieves strong results on the Frama-C benchmark: GPT-OSS~120b converges on 43 of 45 programs (96\%), Qwen-Plus on 40 (89\%), and Claude Haiku~4.5 on 37 (82\%).
Across LF2C-Simple, which compiles the same LF programs into C and runs them through the same pipeline, results are consistent at 82--88\%.
The X509 benchmark is markedly harder: these programs contain pointer-heavy certificate parsing code, and the 33--50\% convergence rate reflects the difficulty of generating sound pointer contracts for low-level memory operations with current LLMs.
Figure~\ref{fig:fm2026_bar} breaks down outcomes (converged, system-only, failed, timeout) by model and strategy on the Frama-C benchmark.

\begin{figure*}[t]
  \includegraphics[width=0.7\linewidth]{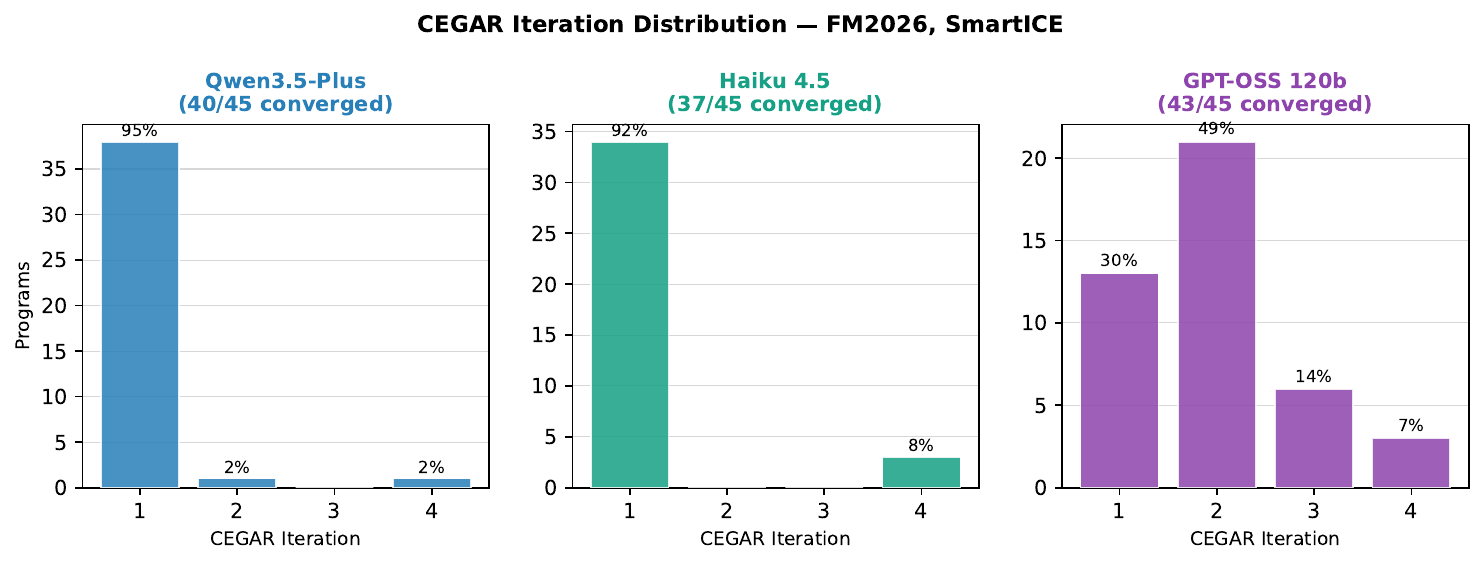}
  \caption{CEGAR iteration distribution on Frama-C benchmark (SmartICE) for three LLM backends.
    Nearly all converged programs require only a single iteration.}
  \label{fig:iter_dist}
\end{figure*}

The Frama-C benchmark results reveal a clear dominance of first-iteration convergence across all three models.
For GPT-OSS~120b, 40 of the 43 converged programs (93\%) require only a single CEGAR iteration; the remaining three converge at iteration~2.
Figure~\ref{fig:iter_dist} shows the full iteration distribution.

\begin{dialogbox}
\paragraph{\textbf{RQ1 --- Verification effectiveness.}}
\ourtool{} achieves 82--96\% on Frama-C benchmark and LF2C-Simple, with a narrow spread across three backends, suggesting the CEGAR+ICE loop is largely model-agnostic for well-structured algorithmic programs.
X509 (33--50\%) is the outlier: pointer-heavy certificate parsing exposes a gap in current LLM contract synthesis for low-level memory code that no backend closes.
\end{dialogbox}

%%%%%%%%%%%%%%%%%%%%%%%%%%%%%%%%%%%%%%%%%%%%%%%%%%%%%%%%%%%%%
\subsection{ICE Learning Ablation (RQ2)}
\label{subsec:ablation}
%%%%%%%%%%%%%%%%%%%%%%%%%%%%%%%%%%%%%%%%%%%%%%%%%%%%%%%%%%%%%

\begin{figure}[t]
  \includegraphics[width=\linewidth]{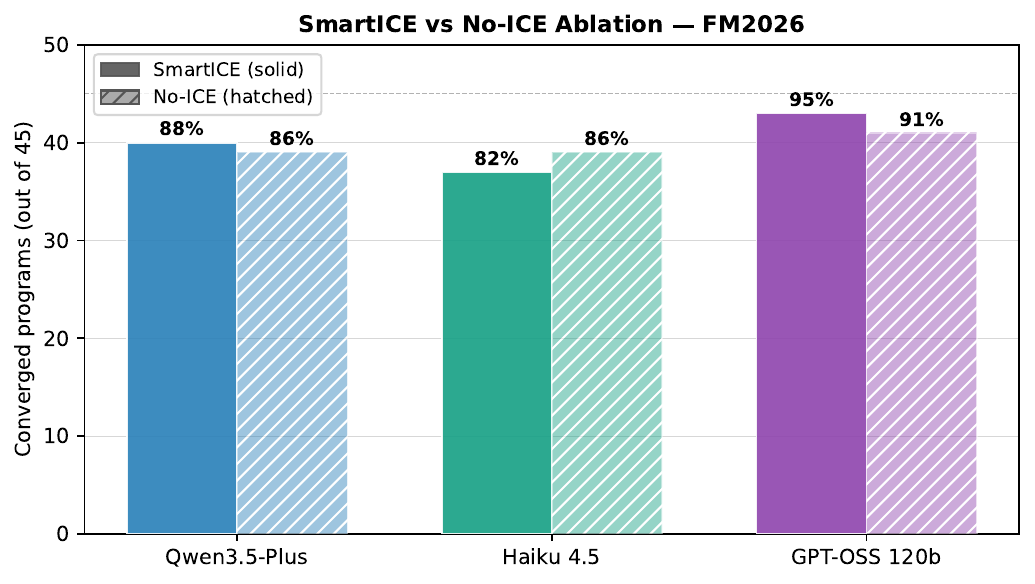}
  \caption{SmartICE vs.\ No-ICE on the Frama-C benchmark for three LLM backends.
    Each pair of bars shows converged program counts with and without ICE learning.}
  \label{fig:ice_ablation}
\end{figure}

To isolate the contribution of SMART ICE learning, we compare SmartICE and No-ICE on the Frama-C benchmark for all three backends.
Figure~\ref{fig:ice_ablation} visualises the comparison.

The ablation reveals a picture more nuanced than a simple verdict on ICE.
GPT-OSS~120b gains the most from structured counterexample feedback (+2 programs, 96\% vs.\ 91\%), and Qwen-Plus benefits similarly (+1 program, 89\% vs.\ 87\%).
Claude Haiku~4.5 tells a different story: it converges on two fewer programs under SmartICE (37/45) than without it (39/45).

The divergence has a traceable cause.
The value of an ICE database depends entirely on the informational quality of its entries: a negative example is useful only if it captures a genuine semantic property of the failed contract, not a surface artefact of how the contract was written.
For Claude Haiku~4.5, two failure modes corrupt this signal.
First, when the model produces a syntactically ill-formed contract---such as using a boolean literal in a context where ESBMC expects a C integer expression---the resulting parse error is stored as a negative example, and the ICE loop constrains subsequent refinement away from that surface pattern.
The model responds by producing other malformed expressions rather than resolving the underlying cause, trapping the loop until the iteration budget expires.
Second, when ESBMC emits counterexamples in a format that ConVer's parser cannot decode, those semantically empty entries populate the E$^-$ database and are presented back to the model as though they carried structural meaning---noise wearing the appearance of signal.
Larger backends, by contrast, generate syntactically well-formed contracts more reliably and elicit richer, more structured ESBMC witnesses, keeping their counterexample databases clean and the feedback loop genuinely informative.
ICE, in this light, functions as a \emph{capability amplifier}: it sharpens refinement precisely where the LLM already generates high-quality artefacts, and offers diminishing---or in the extreme, negative---returns where those artefacts are noisier.

\begin{dialogbox}
\paragraph{\textbf{RQ2 --- ICE learning contribution.}}
SMART ICE learning yields its strongest gains for the highest-capability backend, lifting GPT-OSS~120b by five percentage points (96\% vs.\ 91\%) and Qwen-Plus by two (89\% vs.\ 87\%).
Claude Haiku~4.5, whose smaller capacity leads to more frequent contract syntax errors and less structured ESBMC witnesses, does not benefit: its ICE database accumulates noise rather than signal, and the feedback loop becomes counterproductive.
The finding reframes the role of ICE: rather than a uniform improvement over pure CEGAR, it is a \emph{capability-conditioned amplifier}---most valuable precisely where the LLM is already strong enough to generate artefacts from which structured feedback can be meaningfully extracted.
\end{dialogbox}

%%%%%%%%%%%%%%%%%%%%%%%%%%%%%%%%%%%%%%%%%%%%%%%%%%%%%%%%%%%%%
\subsection{Results on Complex Programs (RQ3)}
%%%%%%%%%%%%%%%%%%%%%%%%%%%%%%%%%%%%%%%%%%%%%%%%%%%%%%%%%%%%%

\subsubsection*{LF-Hard Benchmarks (24 programs)}

\begin{figure}[t]
  \includegraphics[width=\linewidth]{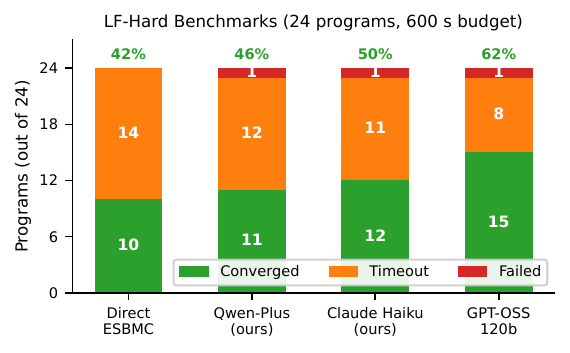}
  \caption{Outcome breakdown on LF-Hard benchmarks (24 programs, 600\,s budget). Direct ESBMC (42\%) is the no-contract baseline. GPT-OSS~120b leads at 62\%, with Claude Haiku~4.5 at 50\% and Qwen-Plus at 46\%, confirming that stronger contract synthesis yields higher convergence.}
  \label{fig:lf_hard}
\end{figure}

Table~\ref{tab:lf_hard_results} shows results on the 24 LF-Hard benchmarks (1{,}998--4{,}328\,LOC each) under the Pre-Abstraction strategy with a 600\,s budget. Figure~\ref{fig:lf_hard} visualises the outcome breakdown.

\begin{table}[t]
  \centering
  \small
  \caption{Pre-Abstraction results on LF-Hard benchmarks (24 programs, 600s budget). ``Direct ESBMC'' verifies without contracts; \ourtool{} columns use Pre-Abstraction. Qwen~=~Qwen-Plus; Claude~=~Haiku~4.5; GPT~=~GPT-OSS~120b.}
  \label{tab:lf_hard_results}
  \begin{tabular}{lcccc}
    \toprule
                & Direct ESBMC & Qwen3.5-Plus       & Haiku4.5              & GPT-OSS \\
    \midrule
    Converged   & 10 (42\%)    & 11 (46\%)  & 12 (50\%)  & \textbf{15 (62\%)} \\
    Timeout     & 14           & 12         & 11                  & 8 \\
    Failed      & 0            & 1          & 1                   & 1 \\
    \bottomrule
  \end{tabular}
\end{table}

Direct ESBMC without contracts verifies 10 of 24 programs within budget, providing a baseline: these are already simplified C translations of LF reactor models that ESBMC can handle monolithically. \ourtool{} improves on this across all three backends: GPT-OSS~120b leads at 15/24 (62\%), followed by Claude Haiku~4.5 at 12/24 (50\%) and Qwen-Plus at 11/24 (46\%). The gains stem from programs such as \texttt{Election}, \texttt{ADASModel}, and \texttt{ProcessMsg} where monolithic ESBMC exploration times out, but contract-based decomposition reduces each function-level SMT instance to a tractable size. GPT-OSS~120b additionally converges on \texttt{Fibonacci} and \texttt{TrainDoorFeedback} where the smaller models time out, reflecting higher contract synthesis quality from the larger model. The ordering of results mirrors the model's capability, confirming that \emph{contract synthesis quality, not ESBMC's solving capacity}, is the binding constraint on these benchmarks.

Most converged programs require no CEGAR refinement (iteration~0); the few that do (e.g., \texttt{Election} at iter=1, \texttt{Fibonacci} at iter=2 for GPT-OSS~120b) converge quickly, indicating the refinement loop is effective when initial contracts are close to correct.

\texttt{ProcessSync} is the single \textsf{FAILED} program across all LLM backends; ESBMC rejects its structure at the system-level check before contract synthesis begins, so it is excluded from the convergence count.

%\paragraph{Soundness note.}
%ESBMC's function-contract mechanism does not support pointer-dereference targets in \texttt{assigns} clauses (e.g., \texttt{assigns *instance\_args}). \ourtool{} strips such clauses before invoking ESBMC, so frame conditions for state modified through pointer parameters are not mechanically checked. All other contract clauses---\texttt{requires}, \texttt{ensures}, and scalar \texttt{assigns}---are verified normally.

%%%%%%%%%%%%%%%%%%%%%%%%%%%%%%%%%%%%%%%%%%%%%%%%%%%%%%%%%%%%%

\subsubsection*{VerifyThis Benchmarks (11 programs)}

%%%%%%%%%%%%%%%%%%%%%%%%%%%
For VerifyThis (11 recursive and loop-invariant programs), we apply the Pre-Abstraction strategy.
Table~\ref{tab:complex_results} summarises the outcomes.

\begin{table}[t]
  \centering
  \small
  \caption{Pre-Abstraction strategy results on VerifyThis (900\,s timeout).
    Qwen = Qwen-Plus; Claude = Haiku~4.5; GPT = GPT-OSS~120b.}
  \label{tab:complex_results}
  \begin{tabular}{lccc}
    \toprule
    Benchmark             & Qwen3.5-Plus        & Haiku4.5              & GPT \\
    \midrule
    VerifyThis (11)       & 6 (55\%)    & \textbf{7 (64\%)}   & 6 (55\%) \\
    \bottomrule
  \end{tabular}
\end{table}

On VerifyThis, Claude Haiku~4.5 achieves 7/11 (64\%), matching the reference result produced by the pipeline authors under manual configuration.
Qwen-Plus and GPT-OSS~120b both reach 6/11 (55\%).
The four programs that fail across all backends --- \texttt{MultCommutative}, \texttt{duplets}, \texttt{prefixsum}, and \texttt{prefixsum\_rec} --- involve complex inter-procedural invariants that exceed the current contract synthesis capability within the iteration budget.

\begin{dialogbox}
\paragraph{\textbf{RQ3 --- Performance on complex programs.}}
On LF-Hard (24 programs), Pre-Abstraction lifts convergence from 42\% (direct ESBMC) to 62\% (GPT-OSS~120b), with results ordering by model capability (GPT-OSS~120b $>$ Claude $>$ Qwen), confirming that contract synthesis quality is the binding constraint.
On VerifyThis (11 programs), Pre-Abstraction achieves 55--64\%, with Claude Haiku~4.5 matching the expert reference at 7/11; the four failing programs involve complex inter-procedural invariants beyond current synthesis capability.
\end{dialogbox}

%%%%%%%%%%%%%%%%%%%%%%%%%%%%%%%%%%%%%%%%%%%%%%%%%%%%%%%%%%%%%
\section{Discussion}
\label{sec:discussion}
%%%%%%%%%%%%%%%%%%%%%%%%%%%%%%%%%%%%%%%%%%%%%%%%%%%%%%%%%%%%%
\paragraph{Comparison with state of the art.}
The most closely related tool is \textsc{PolyVer}~\cite{https://doi.org/10.34727/2025/isbn.978-3-85448-084-6_10}, which also applies LLM-based contract synthesis in a CEGIS--CEGAR loop to verify Lingua Franca programs.
The two systems are complementary rather than directly competing.
\textsc{PolyVer} operates at the \emph{LF source level}: each benchmark comprises 18--206~LOC of LF source and 2--64~LOC of C reaction code, and UCLID5 is used as the system-level model checker, enabling verification of temporal properties expressed as extended state machines and support for polyglot (C and Rust) reactions.
\ourtool{} operates on the \emph{fully translated C output} of \texttt{ESBMC-LF}, where the same benchmarks expand to 1\,998--4\,328~LOC monolithic C files including the complete LF runtime.
This makes \ourtool{}'s verification task substantially harder---state explosion in the translated code is why direct ESBMC fails on 14 of 24 benchmarks---but it also makes \ourtool{} independent of any LF-specific toolchain: it verifies standard C programs without requiring a separate model-checking frontend.
Where \textsc{PolyVer} achieves 22/22 on its benchmark set at the source level, \ourtool{} achieves 15/24 (GPT-OSS~120b) on the translated-C versions of the same programs, overall 16/24 across all run-throughs, a result that is not directly comparable but highlights the additional challenge of verifying runtime-inclusive translations.

\textsc{ACSE}~\cite{lu2026array} is evaluated on the same Frama-C and X.509 benchmarks used in this paper, making it a direct point of comparison.
\textsc{ACSE} generates ACSL pre/post-conditions and assigns clauses via array-carrying symbolic execution, and crucially supports \emph{quantified} loop invariants (e.g., $\forall i.\; a[i] \geq 0$) through its symbolic segment abstraction.
\ourtool{} does not generate quantified invariants: ESBMC's function contract mechanism does not support first-order quantifiers in \texttt{requires}/\texttt{ensures} clauses, so the Frama-C benchmark is restricted to the 45 programs whose properties can be expressed without quantification.
Programs that require quantified array invariants are outside \ourtool{}'s current scope, and \textsc{ACSE} handles these cases where \ourtool{} cannot.
Within the quantifier-free subset, \ourtool{}'s CEGAR loop provides a complementary advantage: rather than inferring contracts analytically in a single pass, it iteratively refines contracts against formal counterexamples, which handles cases where symbolic invariant inference is imprecise or the program is too large for exhaustive symbolic exploration.

\paragraph{Persistent system-only cases.}
Across all four configurations, 4--5~programs reach the system-only outcome:
the system-level ESBMC check passes, but at least one function-level check
fails within the iteration limit.  Inspection reveals that these programs
contain loop-heavy functions for which the LLM consistently produces
contracts that are too strong for the implementation to satisfy, and the 5-iteration budget is insufficient to relax them to a valid specification.
Increasing the iteration limit or adding an automatic loop invariant synthesis step for these functions are promising remedies.

\paragraph{Threats to validity.}
Three threats apply. First, the Frama-C benchmark consists entirely of relatively simple single-assertion programs; generalisation to multi-property, multi-function programs remains to be demonstrated. Second, LLM outputs are stochastic, so results may vary across runs; we ran each configuration once and have not yet conducted repeated-trial experiments. Third, the one known soundness caveat is specific to ESBMC's \texttt{assigns} clause implementation: pointer-dereference targets (e.g., \texttt{assigns *instance\_args}) are not currently supported by ESBMC and are stripped from contracts before invocation, so frame conditions for state modified through pointer parameters are not mechanically verified. This is a limitation of the ESBMC backend, not of the compositional method: all \texttt{requires}, \texttt{ensures}, and scalar \texttt{assigns} clauses are verified normally, and the method's soundness argument holds for the verified portion of each contract. All results reported in the tables are valid under these conditions.

%%%%%%%%%%%%%%%%%%%%%%%%%%%%%%%%%%%%%%%%%%%%%%%%%%%%%%%%%%%%%
\section{Conclusion}
\label{sec:conclusion}
%%%%%%%%%%%%%%%%%%%%%%%%%%%%%%%%%%%%%%%%%%%%%%%%%%%%%%%%%%%%%
We have presented \ourtool{}, a tool for LLM-guided top-down compositional
verification of C programs using ESBMC function contracts.
On the Frama-C benchmark of 45~C programs, \ourtool{} achieves 82--96\%
verification success across three LLM backends (GPT-OSS~120b 96\%, Qwen-Plus 89\%, Claude Haiku~4.5 82\%), with over 93\% of converged programs requiring only a single CEGAR iteration.
On LF2C-Simple (17~programs) results are consistent at 82--88\%, confirming
that the pipeline generalises beyond its primary benchmark.
On the VerifyThis suite of 11~recursive and loop-intensive programs, the
Pre-Abstraction strategy achieves 55--64\%, with Claude Haiku~4.5 matching the expert reference at 7/11.
All three backends achieve competitive results on well-structured programs, while on harder benchmarks (LF-Hard, 46--62\%) results correlate with model capability, confirming that the CEGAR+ICE loop is the primary driver of convergence on tractable programs.

Future work will target two directions: (1) extending the approach to programs
with multiple top-level assertions, and (2) investigating hybrid model selection
strategies that route simple programs to fast, cheap LLMs and complex programs
to stronger reasoning models.

%%%%%%%%%%%%%%%%%%%%%%%%%%%%%%%%%%%%%%%%%%%%%%%%%%%%%%%%%%%%%
\section{Acknowledgements}
%%%%%%%%%%%%%%%%%%%%%%%%%%%%%%%%%%%%%%%%%%%%%%%%%%%%%%%%%%%%%

The authors of this research disclose that generative AI was used in the development of the tooling and in the execution of the methodological evaluation. After using these tool(s)/service(s), the author(s) reviewed and edited the content as needed and take(s) full responsibility for the publication’s content.

%%%%%%%%%%%%%%%%%%%%%%%%%%%%%%%%%%%%%%%%%%%%%%%%%%%%%%%%%%%%%
\section*{Availability}
%%%%%%%%%%%%%%%%%%%%%%%%%%%%%%%%%%%%%%%%%%%%%%%%%%%%%%%%%%%%%

The \ourtool{} tool, all benchmark programs, pre-computed experiment results, and reproduction scripts are openly available as a replication package on Zenodo at \url{https://doi.org/10.5281/zenodo.19249204}. The artifact includes the \ourtool{} source code, ESBMC~v8.1 static binary, all four benchmark suites (Frama-C, X.509, LF2C-Simple, VerifyThis, LF-Hard), and the figures and notebooks used to generate the results in this paper.

\newpage
\bibliographystyle{IEEEtran}
\bibliography{references}

\end{document}